\documentstyle[12pt]{article}
\newcommand\lsim{\mathrel{\rlap{\lower4pt\hbox{\hskip1pt$\sim$}}
        \raise1pt\hbox{$<$}}}
\newcommand\gsim{\mathrel{\rlap{\lower4pt\hbox{\hskip1pt$\sim$}}
        \raise1pt\hbox{$>$}}}

\input epsf

\begin{document}

\title{Flattening the Inflaton's Potential with \\
Quantum Corrections II}
\author{Ewan D. Stewart \\ Research Center for the Early Universe \\
University of Tokyo \\ Tokyo 113, Japan}
\maketitle
\begin{abstract}
In a previous paper I showed that a classical scalar potential with
$ V''/V \sim 1 $ can be sufficiently flattened by quantum corrections
to give rise to slow-roll inflation.
In this paper I give a hybrid inflation implementation of that idea
which can naturally produce a spectral index in the observationally
viable range even for $ V^{1/4} \sim 10^{10} $ to $ 10^{11}\,$GeV.
Although any observationally viable spectral index can be obtained,
the model does predict a distinctive spectral shape.
\end{abstract}
\vspace*{-78ex}
\hspace*{\fill}{RESCEU-9/97}
\thispagestyle{empty}
\setcounter{page}{0}
\newpage
\setcounter{page}{1}

\section{Introduction}
\label{intro}

Slow-roll inflation \cite{KT,Linde} requires an unusually flat scalar
potential.
This is quantified by the conditions
\begin{equation}
\left( \frac{V'}{V} \right)^2 \ll \frac{1}{M_{\rm Pl}^2}
\end{equation}
and
\begin{equation}
\label{src2}
\left| \frac{V''}{V} \right| \ll \frac{1}{M_{\rm Pl}^2}
\end{equation}
where $ M_{\rm Pl} = 1 / \sqrt{8\pi G} $.
The first condition is generally not difficult to achieve; one simply
needs to be sufficiently near an extremum of the potential with
positive potential energy.
In many cases `sufficiently' just means $ \ll M_{\rm Pl} $ from the 
extremum, which is almost difficult to avoid!

However, the second condition presents the most, perhaps only, serious 
obstacle to building a model of inflation \cite{fvi,iss}.
This is because the positive potential energy required for inflation
spontaneously breaks supersymmetry\footnote{
To build a model of inflation, not to mention the rest of particle
physics, in the absence of supersymmetry seems hopeless.
For an introduction to supersymmetry and supergravity, see
Ref.~\cite{susy}.},
generically inducing
\begin{equation}
\left| \frac{V''}{V} \right| \gsim \frac{1}{M_{\rm Pl}^2}
\end{equation}
for all scalar fields \cite{Dine,fvi,iss}, and in particular for the
inflaton \cite{fvi,iss}.
The $\sim$ holds when the field has no supersymmetric mass and the
supersymmetry breaking is only communicated via gravitational strength
interactions.

Many models of inflation are built ignoring gravitational strength
interactions, and so are implicitly setting $ M_{\rm Pl} = \infty $.
Clearly one cannot achieve Eq.~(\ref{src2}) in this context.
Models of inflation built in the context of supergravity
have traditionally resolved this problem by fine tuning, either
explicit or implicit.
Only recently have there been any plausible proposals for solving
this problem.

The first\footnote{
Natural inflation \cite{nat} naturally achieves a small $V''$ by
assuming an approximate global $U(1)$ symmetry, but does {\em not\/}
naturally satisfy Eq.~(\ref{src2}) because $V$ vanishes in the limit
where the symmetry is exact.
A hybrid natural inflation model might avoid this problem though,
as was noted in Ref.~\cite{fip}.}
was given in Refs.~\cite{fvi,iss}.
It employed forms for the Kahler potential that had been derived from
weakly coupled heterotic string theory, in combination with a subset
of the modular symmetries, to cancel the supergravity corrections.
This method should also work in appropriate limits of M-theory.
It requires an inflationary energy density well above the vacuum
supersymmetry breaking scale, $ V^{1/4} \gg M_{\rm s} $;
$ M_{\rm s} \sim 10^{10} $ to $ 10^{11}\,$GeV in our vacuum for
gravity mediated supersymmetry breaking.
A natural implementation of this method would be to have the
non-perturbative physics that leads to gaugino condensation at a scale 
$ \Lambda_{\rm gc} \sim 10^{13}\,$GeV in our vacuum generate the
inflationary potential at a similar energy scale
$ V^{1/4} \sim \Lambda_{\rm gc} $.
It should be possible to stabilise the moduli during inflation by the
same method that stabilises them in our vacuum, though the transition
from inflation to vacuum could be dangerous if the end of inflation is 
not sufficiently smooth.
A simple model of inflation with the appropriate energy scale and a
smooth end to inflation was given in Ref.~\cite{mut}.
This method has the practical disadvantage that it requires detailed
control of the effective supergravity theory, but otherwise remains
very promising.

It was also noted in Ref.~\cite{iss} that the supergravity corrections
could be avoided if the inflationary potential energy was dominated by
the $D$-term, and a hybrid inflation \cite{hybrid,fvi} implementation
of that idea was constructed with a Fayet-Iliopoulos term dominating
the energy density.
The main problem with this method is to obtain a Fayet-Iliopoulos
term at a low enough energy scale to obtain the COBE normalisation for
the density perturbations, but at a scale higher than the $F$-term
supersymmetry breaking to avoid the $F$-term supersymmetry breaking
inducing too large a mass for the inflaton.
The stabilisation of the moduli, the dilaton in the case of the
heterotic string, may also be a serious problem.
A Fayet-Iliopoulos term does arise in many compactifications of the
heterotic string \cite{FI}, but its scale,
typically $ V_{\rm FI}^{1/4} \sim 10^{17} $ to $ 10^{18}\,$GeV,
a conservative lower bound being
$ V_{\rm FI}^{1/4} > 2 \times 10^{16}\,$GeV,
is at the limit of being too high for any type of inflation to meet
the COBE constraint \cite{iss}, and in particular is always too high
if the slope of the inflaton's potential is dominated by the loop
correction as in Ref.~\cite{Dvali}, in which case the COBE
normalisation requires
$ V_{\rm infl}^{1/4} \sim 5 \times 10^{15}\,$GeV.
However, in M-theory scales are more flexible \cite{Witten}, and so it
may be possible to obtain a Fayet-Iliopoulos term at a sufficiently
low scale.
There still remains the problem of stabilising the moduli.
One possibility would be if either the gauge coupling, or the
Fayet-Iliopoulos term itself, had a large non-perturbative dependence
on the moduli.
Whether this can be achieved at the same time as having a sufficiently 
low energy scale for the Fayet-Iliopoulos term remains to be seen.
Another possibility would be to generate the Fayet-Iliopoulos term by
field theory methods at a lower energy scale, but as this probably
requires $F$-term supersymmetry breaking it may be difficult to avoid
the inflaton getting too large a mass from this source.
Especially in this latter case, one might also have to consider the
inflaton dependence of the gauge coupling and Fayet-Iliopoulos term,
which could provide an effective mass for the inflaton. 

The method of \cite{Murayama} uses a global Heisenberg symmetry,
which has been derived from string theory at tree and one loop level,
to cancel the supergravity corrections.
The stabilisation of the modulus that forms an integral part of the
mechanism is a serious problem for this method.
This is the only method that could conceivably implement the naively
popular $\phi^n$ chaotic inflation models.

The method of \cite{Ross} is somewhat complex and requires specific
couplings of a Goldstone boson to the inflaton, but is an interesting
possibility.

In this paper I will use the method of Ref.~\cite{fip}.
It has the advantages that it does not require any special features of
the high energy theory, and can be implemented at the scale
$ V^{1/4} \sim 10^{10} $ to $ 10^{11}\,$GeV where the moduli are
already stabilised.

\begin{figure}
\epsfxsize=414pt
\epsfbox{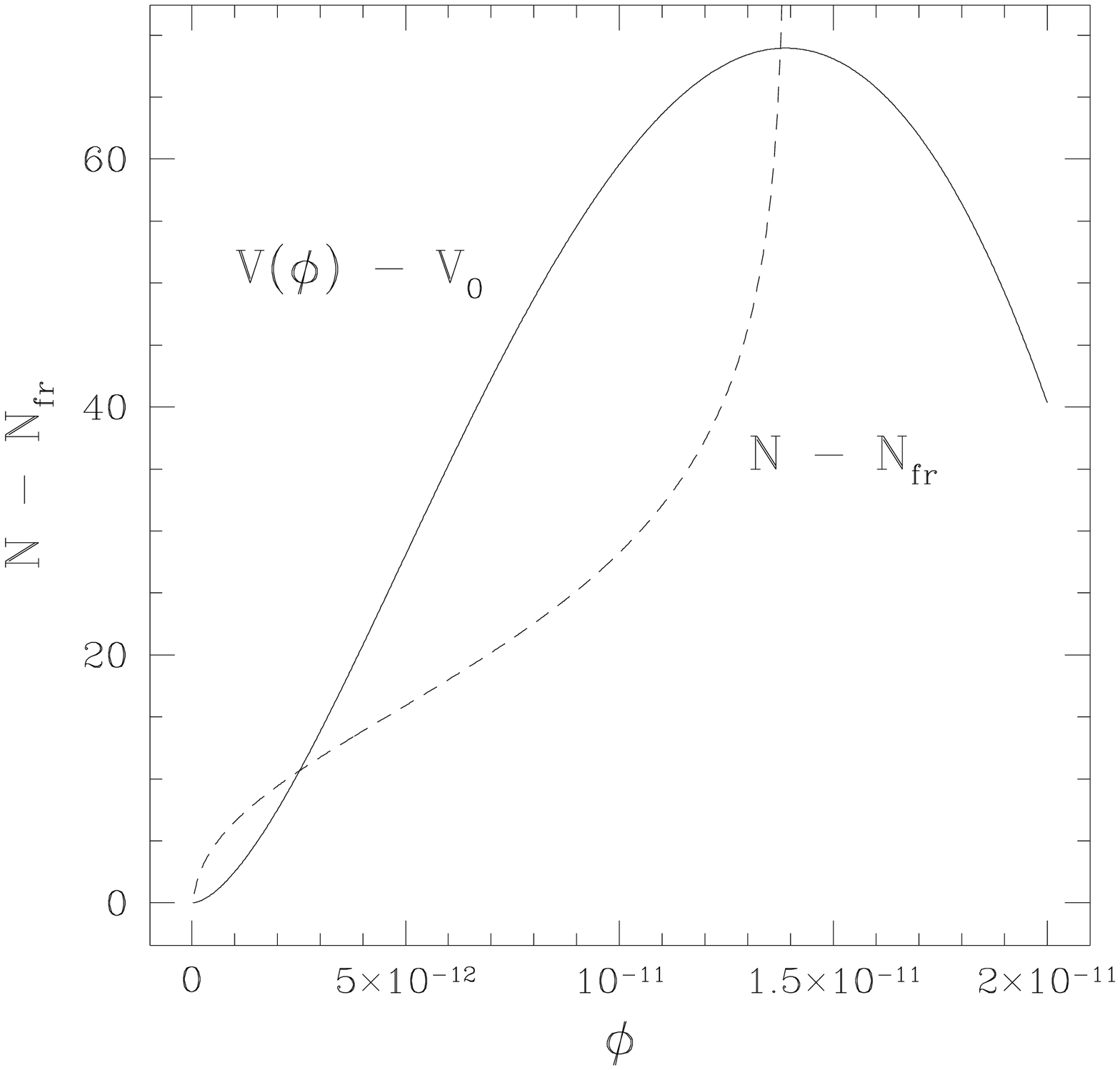}
\caption{
The potential Eq.~(4), with $f$ given by Eqs.~(16) and~(18), for
$ f(0) = 1 $ and $ A = 0.5 $.
The potential energy at $\phi=0$ is subtracted and the normalisation is
arbitrary.
The number of $e$-folds of inflation until the inflaton starts to fast 
roll towards $\phi=0$ is plotted on the same graph.
The smallest and largest observable scales might be expected to
correspond to $ N - N_{\rm fr} \sim 10 $ and 20, respectively, though
there is considerable uncertainty in this estimate.
\label{potfig}
}
\end{figure}

I will rely heavily on Ref.~\cite{fip}.
The reader would benefit from reading it first.
From now on I set $ M_{\rm Pl} = 1 $.
The potential of the model (see Figure~\ref{potfig}) is \cite{fip}
\begin{equation}
\label{pot}
V(\phi) = V_0 \left[ 1 - \frac{1}{2} f(\epsilon\ln\phi) \phi^2
+ \ldots \right]
\end{equation}
with $ f(0) \sim 1 $, the minimum value expected in a generic
supergravity theory, and $\epsilon \ll 1 $.
The function $f$ is determined by the renormalisation group running of
the mass of the inflaton $\phi$.
The potential is assumed to have a maximum at $ \phi = \phi_* $,
with $ V_0^{1/2} \ll \phi_* \ll 1 $.
We then have $ f_* = {\cal O}(\epsilon) $, allowing slow roll inflation 
near $\phi_*$ \cite{fip}.
We are free to choose $ \phi_* = e^{-1/\epsilon} $, so that
$ f_* = f(-1) $.

The most natural scale for the potential is $ V_0 \sim M_{\rm s}^4 $,
where $M_{\rm s}$ is the supersymmetry breaking scale in our vacuum.
I will take $ M_{\rm s} \sim 10^{-8} $ corresponding to gravity
mediated supersymmetry breaking.
In this case the COBE normalisation gives $ \phi_* \sim 10^{-11} $ and 
so $ \epsilon \simeq 0.04 $ \cite{fip}.
In Section~\ref{renorm}, $\epsilon$ will be derived from a gauge
coupling of strength similar to that of the GUT gauge coupling
inferred from the LEP data, $\alpha_{\rm GUT} \sim 0.04$.
It is worth noting that the fact that $M_{\rm s}$ is so small is
crucial to our being able to achieve slow roll inflation without fine
tuning.

\section{The renormalisation group}
\label{renorm}

In this section I will give a rough derivation of the function
$f(\epsilon\ln\phi)$.
The inflaton $\phi$ gives masses proportional to its expectation value 
to the fields to which it couples, and so $\phi$ acts like an infra-red 
cutoff to the renormalisation.
I will assume that $\phi$ is charged under some asymptotically free
gauge group and for simplicity neglect the Yukawa couplings.
Defining
\begin{equation}
x \equiv \epsilon \ln \phi
\end{equation}
so that $ x_* = -1 $,
the renormalisation group equation \cite{susy} for $\phi$'s mass becomes
\begin{equation}
\frac{dm_\phi^2}{dx} = - \frac{ 2 c \alpha }{ \pi \epsilon }
\tilde{m}^2
\end{equation}
where $c$ is the quadratic Casimir invariant of $\phi$'s
representation under the gauge group.
For example, $ c = 3/4 $ for a fundamental representation of SU(2) and
$ c = 4/3 $ for that of SU(3).
$ \alpha = g_{\rm gauge}^2 / 4 \pi $, where $g_{\rm gauge}$ is the
gauge coupling.
$\tilde{m}$ is the gaugino mass.
The renormalisation of $\alpha$ is given by
\begin{equation}
\frac{d\alpha}{dx} = - \frac{ b \alpha^2 }{ 2 \pi \epsilon }
\end{equation}
where, for example, $ b = 3 N_{\rm c} - N_{\rm f} $ for an
SU($N_{\rm c}$) gauge group with $N_{\rm f}$ pairs of fundamentals and
antifundamentals.
$b>0$ corresponds to an asymptotically free gauge group.
The renormalisation of $ \tilde{m} $ is given by
\begin{equation}
\tilde{m}(x) = \frac{\alpha(x) \tilde{m}(0)}{\alpha(0)} \,.
\end{equation}
Integrating these equations gives
\begin{equation}
m_\phi^2(x) = - m_\phi^2(0) \left[ A
\left( \frac{1}{ 1 + \frac{b}{2\pi} \frac{\alpha(0)}{\epsilon} x }
\right)^2 - \left( A + 1 \right) \right]
\end{equation}
where
\begin{equation}
A = \frac{ 2c }{ b } \left( \frac{ \tilde{m}^2(0) }{ - m_\phi^2(0) }
\right) \,.
\end{equation}
To lowest order, $\epsilon$ is defined by $ m_\phi^2(-1) = 0 $, which
gives
\begin{equation}
\epsilon = \alpha(0) \frac{b}{2\pi}
\left[ A + 1 + \sqrt{A(A+1)} \right] \,.
\end{equation}
Therefore
\begin{equation}
\label{mphi}
m_\phi^2(x) = - (A+1) m_\phi^2(0)
\left[ \left( \frac{y_\infty}{y_\infty+1+x} \right)^2 - 1 \right]
\end{equation}
where
\begin{equation}
\label{yinf}
y_\infty = A + \sqrt{ A ( A + 1 ) } \,.
\end{equation}
The potential, Eq.~(\ref{pot}), is then
\begin{equation}
V(\phi) = V_0 + \frac{1}{2} m_\phi^2(\epsilon\ln\phi) \phi^2 + \ldots
\end{equation}
with $ m_\phi^2(\epsilon\ln\phi) $ given by Eq.~(\ref{mphi})
and $ m_\phi^2(0) \sim - V_0 $.

\section{Initial conditions}

In Ref.~\cite{fip}, I said that the above model with the slow roll
inflation occuring as $\phi$ rolls from the maximum at
$ \phi = \phi_* $ towards the false vacuum at $\phi=0$, had
problematic initial conditions.
Here, I argue to the contrary that this model has very natural initial
conditions if we choose the vacuum at $ \phi > \phi_* $ to be another
false vacuum.
Then the inflaton $\phi$ will be easily trapped in the large false
vacuum potential well at $ \phi > \phi_* $, giving rise to eternal
\cite{Linde} old inflation.
There will be an extremely small, but {\em finite}, chance that
quantum fluctuations will kick $\phi$ to the top of the barrier at
$ \phi = \phi_* $, where once again eternal inflation occurs.
The chance that this will happen is extremely small, but, because it
is finite, it will be more than compensated for by the two lots of
eternal inflation.
Thus the initial conditions for the slow roll inflation,
$ \phi = \phi_* $, are naturally obtained.

\section{Slow roll inflation}

As we are focussing on the case where the slow roll inflation
occurs as $\phi$ rolls off the maximum at $ \phi = \phi_* $ towards
the false vacuum at $\phi=0$, it will be convenient to define
\begin{equation}
y \equiv \epsilon \ln \left( \frac{\phi_*}{\phi} \right) = - 1 - x \,,
\end{equation}
so that $ y_* = 0 $ and $ y > 0 $ during the slow roll inflation, and
\begin{equation}
g(y) \equiv - f(x) - \frac{\epsilon}{2} f'(x) \,,
\end{equation}
so that $ g_* = 0 $, $ g > 0 $ during the slow roll inflation, and
$ g_*' = f_*' + {\cal O}(\epsilon) $.
Eq.~(19) of Ref.~\cite{fip} then becomes
\begin{equation}
\label{tauint}
\tau = \epsilon H t
= \frac{2}{3} \int \frac{ dy }{ 1 - \sqrt{ 1 - 4g/3 } } \,.
\end{equation}
Following Section~\ref{renorm}, we take as an example
\begin{equation}
g(y) = a \left[ \left( \frac{y_\infty}{y_\infty-y} \right)^2 - 1 \right]
\end{equation}
with
\begin{equation}
a = ( 1 + A ) f(0) + {\cal O}(\epsilon)
\end{equation}
and $y_\infty$ given by Eq.~(\ref{yinf}).
Then
\begin{equation}
\label{gp}
g_*' = \frac{2a}{y_\infty} \,.
\end{equation}
With this choice for $g$, we can integrate Eq.~(\ref{tauint}) to give
\begin{equation}
\label{tau}
\tau = \frac{1}{g_*'} \left\{ \ln y
+ \left( 1 - \frac{y}{y_\infty} \right)
 \left( 1 + \sqrt{ 1 - \frac{4g}{3} } \right)
- \ln \left[ 1 + \left( 1 - \frac{y}{y_\infty} \right)
 \sqrt{ 1 - \frac{4g}{3} } \right]
- 2 + \ln 2 \right\} \,.
\end{equation}
This is only valid for $ g < g_{\rm fr} = 3/4 $ or
\begin{equation}
\label{yfr}
y < y_{\rm fr}
= y_\infty \left( 1 - \frac{ 1 }{ \sqrt{ 1 + \frac{3}{4a} } } \right)
\end{equation}
or
\begin{equation}
\label{tfr}
\tau < \tau_{\rm fr} = \frac{1}{g_*'} \left( \ln y_{\rm fr}
+ \frac{ 1 }{ \sqrt{ 1 + \frac{3}{4a} } } - 2 + \ln 2 \right) \,.
\end{equation}
For $ \tau > \tau_{\rm fr} $, $\phi$ fast rolls towards $\phi=0$.
I will neglect the small number of $e$-folds of inflation during this
fast rolling stage.
During slow roll, $ y \ll 1 $ and $ g \ll 1 $, and so \cite{fip}
\begin{equation}
\label{tsr}
\tau \simeq \tau_{\rm sr} = \frac{1}{g_*'} \ln y \,.
\end{equation}

\section{Ending inflation}
\label{end}

To end inflation, one must use a hybrid inflation type mechanism
\cite{hybrid,mut,inv} to exit from the false vacuum at some critical
value $ \phi_{\rm c} < \phi_* $.

A standard hybrid inflation exit from the false vacuum \cite{hybrid},
with the new field $\psi$ also having a mass squared
$ m_\psi^2 \sim V_0 $ \cite{Lisa}, and with a coupling
$ \lambda \lsim 1 $ between the fields,
\begin{equation} 
V(\phi, \psi) = V_0
+ \frac{1}{2} \left[ g(y) + {\cal O}(\epsilon) \right] V_0 \phi^2
- \frac{1}{2} m_\psi^2 \psi^2 + \frac{1}{2} \lambda^2 \phi^2 \psi^2
+ \ldots 
\end{equation}
would give $ \phi_{\rm c} = m_\psi / \lambda \gsim 10^{-16} $,
or $ y_{\rm c} \lsim 0.46 $.
Typically, $ g_{\rm c} > 3/4 $, in which case $\phi$ will be fast
rolling at $ \phi = \phi_{\rm c} $ making the precise value of
$\phi_{\rm c}$ not very important.

Such a hybrid inflation exit from the false vacuum may lead to a large
spike in the spectrum \cite{Lisa}, which could be dangerous.
In our model, contrary to that of Ref.~\cite{Lisa}, the inflaton is
typically fast rolling at $ \phi = \phi_{\rm c} $ which may make the
spike less dangerous \cite{Lisa}.
If dangerous, such a spike could be avoided by a mutated hybrid
inflation type exit from the false vacuum \cite{mut,Juan,inv}.
For example, one could naturally have an additional term
$ \sim V_0 \phi \psi $ in the potential which would avoid the spike
but leave $ \phi_{\rm c} $ essentially unchanged.
As the gauge coupling is getting stronger as $\phi$ is decreasing, it
is also conceivable that the exit from the false vacuum might be
controlled by strong coupling effects.

Finally, there will be a stage of non-slow roll inflation \cite{Lisa}
in which $\psi$ rolls from $ \psi = \psi_{\rm init} \sim 10^{-16} $ to
our vacuum at $ \psi = \psi_{\rm vac} \sim 1 $.
It will last for
\begin{eqnarray}
N_{\psi} & = & \frac{V_0}{2 m_\psi^2}
\left( 1 + \sqrt{ 1 + \frac{4 m_\psi^2}{3V_0} } \right)
\ln \frac{\psi_{\rm vac}}{\psi_{\rm init}} \\
\label{Npsi}
& \sim & \frac{37 V_0}{2 m_\psi^2}
\left( 1 + \sqrt{ 1 + \frac{4 m_\psi^2}{3V_0} } \right)
\end{eqnarray}
$e$-folds.

For $ V_0 \sim 10^{-32} $, and assuming that thermal inflation
\cite{thermal} occurs after this inflation, observable scales would
leave the horizon between 20 and 30 $e$-folds before the end of this
inflation \cite{David}.
If reheating was rapid and there was no thermal inflation, this would
be increased to between 40 and 50 $e$-folds at maximum.
For $ m_\psi^2 = V_0 $, Eq.~(\ref{Npsi}) gives $ N_\psi = 47 $, 
so that even in the latter case observable scales would leave
the horizon during this non-slow roll stage of inflation leading to an
unacceptable spectral index.
To get observable scales safely leaving the horizon during the slow
roll inflation, say at least between 10 and 20 $e$-folds before $\phi$
reached $ \phi = \phi_{\rm c} $, would require $ N_\psi \lsim 10 $ and 
so $ m_\psi^2 \gsim 8 V_0 $ in the case with thermal inflation, and
$ N_\psi \lsim 30 $ and so $ m_\psi^2 \gsim 1.7 V_0 $ in the case
without.
The latter requirement is clearly not very severe, and even the former
is not unreasonable. For example,
$ V(\psi) = \frac{1}{2} V_0
\left[ 1 + \cos \left( 2 \pi \psi \right) \right] $,
which has a period of one, gives $ m_\psi^2 = 2 \pi^2 V_0 $.
However, I do not find this feature of the model entirely
satisfactory in the case with thermal inflation.

\section{The spectral index}

To lowest order in the slow roll approximation, the spectral index is
\cite{fip}
\begin{equation}
n = 1 + 2 \frac{V''}{V}
= 1 + 2 g'_* \left( y - \epsilon \right) \,.
\end{equation}
Eqs.~(\ref{gp}), (\ref{yfr}), (\ref{tfr}) and (\ref{tsr}) then give 
\begin{equation}
\label{n}
n = 1 - 2 \epsilon g'_*
+ 8 a \left( 1 - \frac{ 1 }{ \sqrt{ 1 + \frac{3}{4a} } } \right)
\exp \left[ - \epsilon g'_* \left( N - N_{\rm fr} \right) - 2
+ \frac{ 1 }{ \sqrt{ 1 + \frac{3}{4a} } } \right]
\end{equation}
where $ N - N_{\rm fr} = ( \tau_{\rm fr} - \tau ) / \epsilon $ is the
number of $e$-folds of inflation from the scale $ k \propto e^{-N} $
leaving the horizon to $ \phi = \phi_{\rm fr} $.
From the discussion of Section~\ref{end}, one might expect observable
scales to leave the horizon during the interval
$ N - N_{\rm fr} \sim 10 $ to 20. 
I have been assuming that $ \phi_{\rm c} < \phi_{\rm fr} $, as will
usually be the case.
If $ \phi_{\rm c} > \phi_{\rm fr} $, inflation may end slightly earlier
so that one should take a slightly larger value of $ N - N_{\rm fr} $.  

\begin{figure}
\epsfxsize=414pt
\epsfbox{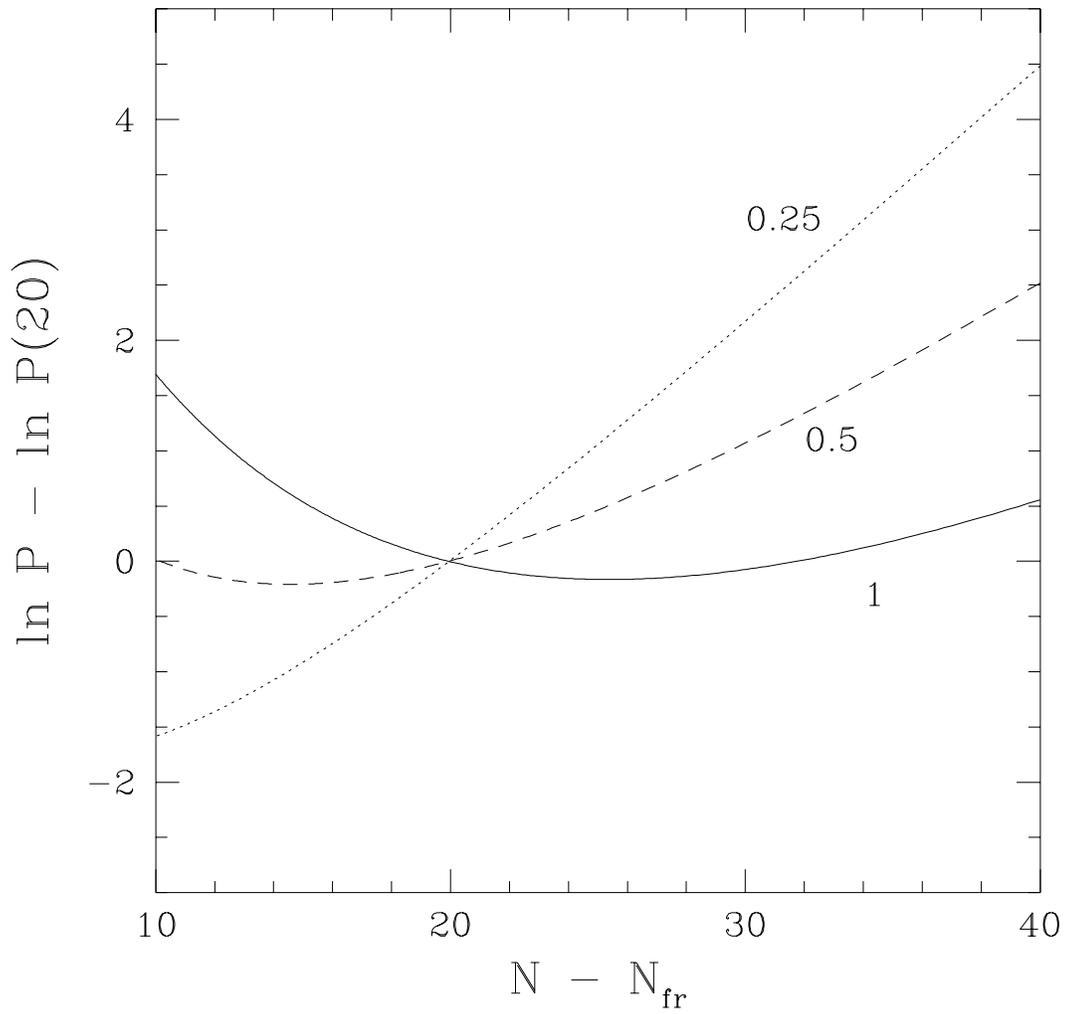}
\caption{
The spectra for $f(0)=1$ and the displayed values of $A$.
\label{specfig}
}
\end{figure}

\begin{figure}
\epsfxsize=414pt
\epsfbox{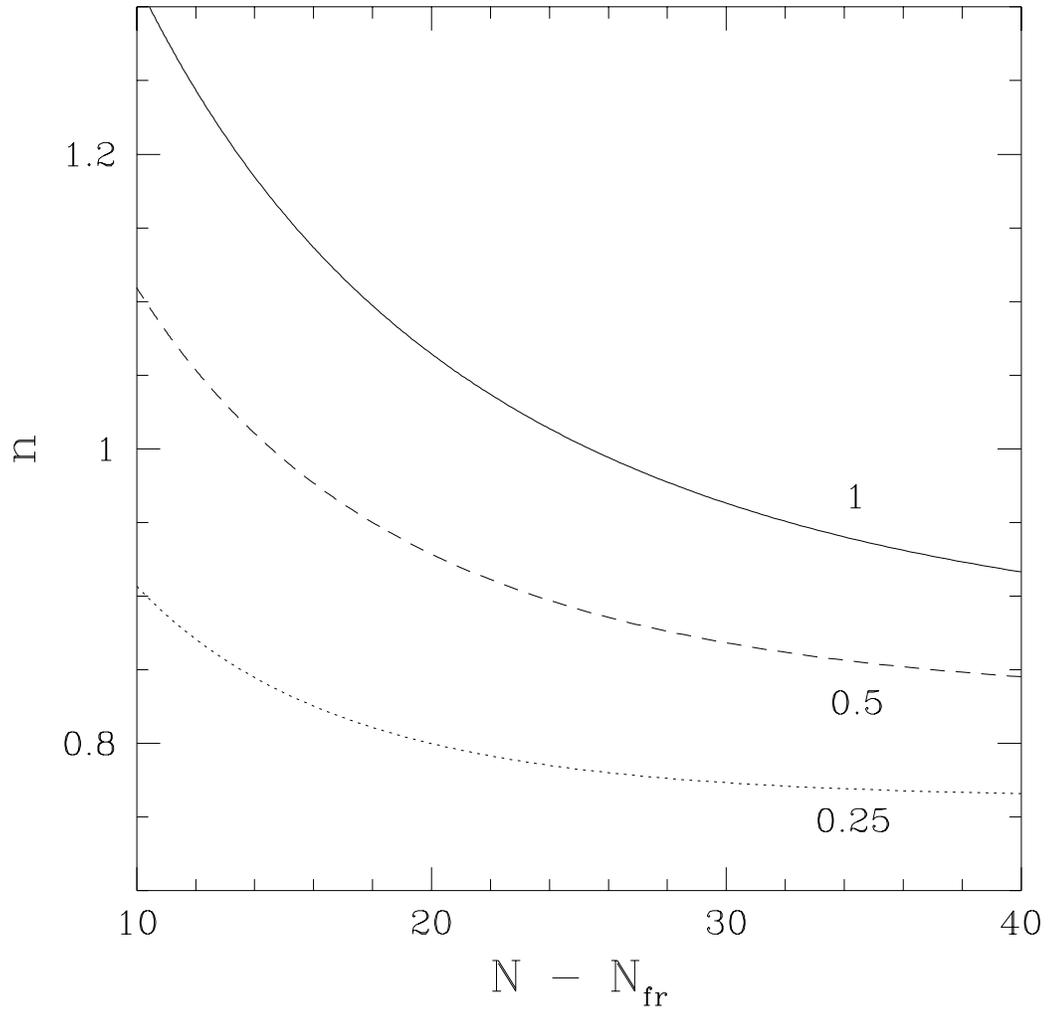}
\caption{
The spectral indices for $f(0)=1$ and the displayed values of $A$.
\label{indfig}
}
\end{figure}

A more accurate but implicit formula is given by using
\cite{2nd}
\begin{eqnarray}
n & = & 1 + 2 \frac{V''}{V} + 2.13 \frac{V'V'''}{V^2}
+ \frac{2}{3} \left( \frac{V''}{V} \right)^2 \\
& = & 1 + 2 \left( g - \epsilon g' \right)
- 2.13 \epsilon g \left( g' - \epsilon g'' \right)
+ \frac{2}{3} \left( g - \epsilon g' \right)^2
\end{eqnarray}
and Eq.~(\ref{tau}) instead of Eq.~(\ref{tsr}).
Some example spectra and spectral indices are plotted in
Figures~\ref{specfig} and~\ref{indfig} respectively.

\section{Conclusions}

In a previous paper \cite{fip} I described how to construct a
potential flat enough for slow roll inflation without fine tuning and
without making any assumptions about the high energy theory.
In this paper I have emphasised an implementation of that idea which
can naturally produce an observationally viable spectral index
even for $ V^{1/4} \sim 10^{10} $ to $ 10^{11}\,$GeV.
Although any observationally viable spectral index can be obtained,
the model does predict a distinctive shape to the spectrum,
given to lowest order in the slow roll approximation by
\begin{equation}
P(k) = Q k^{-2\nu} \exp \left( \sigma k^\nu \right) \,,
\end{equation}
or in terms of the spectral index
\begin{equation}
n(k) = 1 - 2 \nu + \sigma \nu k^\nu \,,
\end{equation}
which may allow it to be distinguished from other models of inflation 
by sufficiently precise observations.
The extra power on small scales, for a given tilt on larger scales,
may be helpful for mixed dark matter scenarios of structure formation.

Finally, I will compare the model with that of Randall, Soljacic and
Guth (RSG) \cite{Lisa}.
The models share some of the same motivations and features, such as to 
use the flat directions characteristic of supersymmetric and string
theories, lifted by supersymmetry breaking at a scale
$ M_{\rm s} \sim 10^{10} $ to $ 10^{11}\,$GeV, to drive inflation.
Both also use a hybrid inflation type mechanism to exit from the false
vacuum.
However, they also have important differences:
\begin{enumerate}
\item
RSG fine tune the inflaton's mass while our model uses the method of
Ref.~\cite{fip}, \mbox{i.e.} the renormalisation of the inflaton's mass
that would be present anyway, to obtain a mass small enough for slow
roll inflation without fine tuning.
\item
The numbers tend to work out better here, \mbox{e.g.} the coupling
$\lambda$ involved in the hybrid inflation mechanism does not need to
be small; our small parameter $\epsilon$ is derived from a gauge
coupling of the order of the GUT gauge coupling inferred from the LEP
data.
The only parameter we may have to tune slightly is the mass of the
other field involved in the hybrid inflation mechanism, but even this
was to a lesser extent than in RSG.
\item
The possible spike in the spectrum produced by the hybrid inflation
mechanism is probably less dangerous here because the inflaton will
typically be fast rolling at this point.
\item
The spectral index in RSG is $ n > 1 $, while our model can give
$ n < 1 $ which may be preferred by observations.
\item
The spectrum has a significant bend. It is not clear whether this is
an advantage or not, but is at least observationally interesting.
\end{enumerate}

\subsection*{Acknowledgements}
I thank David Lyth for helpful discussions.
I am supported by a JSPS Fellowship at RESCEU, and my work is
supported by Monbusho Grant-in-Aid for JSPS Fellows No.\ 96184.

\end{document}